\documentclass[letterpaper,twocolumn,10pt]{article}
\usepackage{usenix-2020-09}

\usepackage{tikz}
\usepackage{amsmath}
\usepackage{xspace}
\usepackage{enumitem}

\usepackage{booktabs}
\usepackage{listings}
\usepackage{cleveref}
\usepackage{algorithm}
\usepackage[noend]{algpseudocode}
\usepackage{amssymb}
\usepackage{multirow}
\usepackage{makecell}
\usepackage{titlesec}
\usepackage{enumitem}
\usepackage{caption}

\usepackage{amsthm}

\newtheorem{definition}{Definition}

\newtheorem{theorem}{Theorem}

\newtheorem{remark}{Remark}
\newcommand{\sys}{VTC\xspace}
\newcommand{\VTOG}{{\rm VTOG}}
\newcommand{\PTG}{{\rm PTG}}

\begin{document}

\date{}

\title{\Large \bf \sys: DNN Compilation with Virtual Tensors for Data Movement Elimination}

\author{
{\rm Muyan Hu}$^1$\thanks{Part of the work done during an internship at Microsoft.} \hspace{1em}
{\rm Ahan Gupta}$^1$ \hspace{1em}
{\rm Jiachen Yuan}$^1$ \hspace{1em}
{\rm Vima Gupta}$^2$ \hspace{1em}
{\rm Taeksang Kim}$^1$ \hspace{1em}
{\rm Xin Xu}$^1$ \\
{\rm Janardhan Kulkarni}$^3$ \hspace{1.2em}
{\rm Ofer Dekel}$^3$ \hspace{1.2em}
{\rm Vikram Adve}$^1$ \hspace{1.2em}
{\rm Charith Mendis}$^1$ \\[.3em]
$^1$University of Illinois Urbana-Champaign\hspace{1.2em} $^2$Georgia Institute of Technology\hspace{1.2em}$^3$Microsoft
}

\maketitle

\begin{abstract}
With the widening gap between compute and memory operation latencies, data movement optimizations have become increasingly important for DNN compilation.
Current optimizations such as layout transformations and operator fusion only target a subset of tensor operators and consequently miss important opportunities for reducing data movement in contemporary DNN workloads, including large language models.

We introduce \sys, a novel tensor compilation framework that for the first time eliminates all unnecessary data movement by targeting the full spectrum of data movement operators.
\sys proposes the concept of \emph{virtual tensors} to track data movement between compute operators via index mappings rather than expensive physical data transfers to and from global memory, which can seamlessly interoperate with existing computation kernels and handle arbitrary tensor operator compositions.
We also introduce a novel data movement elimination algorithm to automatically identify a profitable virtual tensor creation strategy.
Evaluation on a variety of DNNs shows that \sys can outperform existing ML compilers by up to $1.93\times$ ($1.28\times$ on average) on NVIDIA GPUs with up to 60\% (17.5\% on average) inference memory savings.

\end{abstract}

\section{Introduction}

\noindent Deep neural network (DNN) workloads have gained popularity in recent years. Usually, DNN models are expressed as tensor (i.e. $n-$dimensional array) based computations.
Consequently, DNN developers express these computations using tensor programming languages such as TensorFlow~\cite{abadi2016tensorflow}, JAX~\cite{jax2018github}, and PyTorch~\cite{ansel2024pytorch}, and then utilize tensor compilers like XLA~\cite{xla}, TorchInductor~\cite{ansel2024pytorch}, and TVM~\cite{chen2018tvm} to generate highly performant executables targeting various hardware devices.
These compilers use a multi-stage pipeline similar to general-purpose compilers~\cite{auslander1982overview}.
During the \emph{compiler frontend}, computations expressed in tensor programming languages are transformed into a compiler intermediate representation (IR), known as \emph{computation graph}.
Nodes of these graphs represent tensor operators (e.g. matrix multiplications, convolutions), and each edge represents a tensor flowing from the output of a producer node to the input of a consumer node.
The \emph{compiler middle-end} performs graph-level transformations (e.g. operator fusion, tiling) to optimize the computation graph.
Finally, the \emph{compiler backend} maps the computation graph onto a set of kernels, each of which is a program fragment written in single-program-multiple-data (SPMD) fashion on modern hardware accelerators.

To keep up with the computational demands of DNN models, specialized hardware accelerators, such as NVIDIA Tensor Cores, have been introduced.
Each new generation of these accelerators has consistently improved the compute capacity to provide blazing speeds for compute instructions such as matrix multiplications (e.g., nearly 1 PFLOPS half precision on NVIDIA H100).
Most tensor compilers already aim to maximize utilization on such hardware using dedicated compiler backends (e.g. Triton~\cite{tillet2019triton} for GPUs), achieving low latency and high throughput for compute operations.

\begin{figure}
    \centering
    \includegraphics[width=0.75\linewidth]{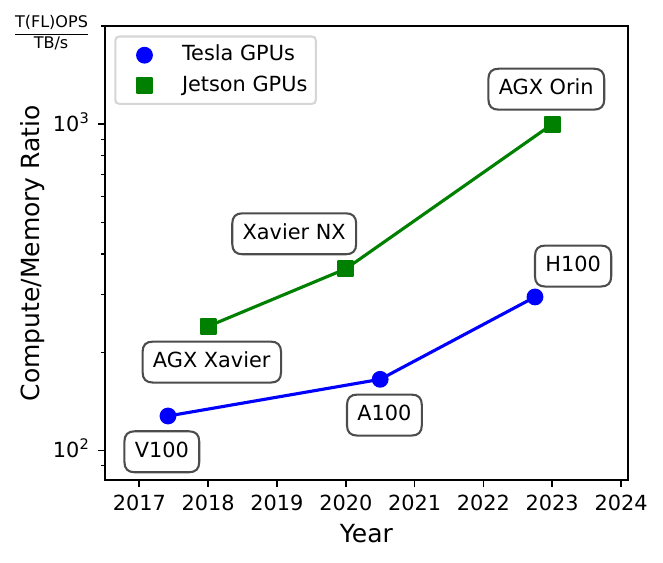}

    \caption{Trend of compute/memory ratio for NVIDIA GPUs over time. The ratio is calculated by dividing the computation power (half-precision performance in TFLOPS for Tesla GPUs and INT8 performance in TOPS for Jetson Edge GPUs) by the peak GPU memory bandwidth (in TB/s). }
    \label{fig:ratio}
\end{figure}

\begin{figure}[t!]
    \centering
    \includegraphics[width=\linewidth]{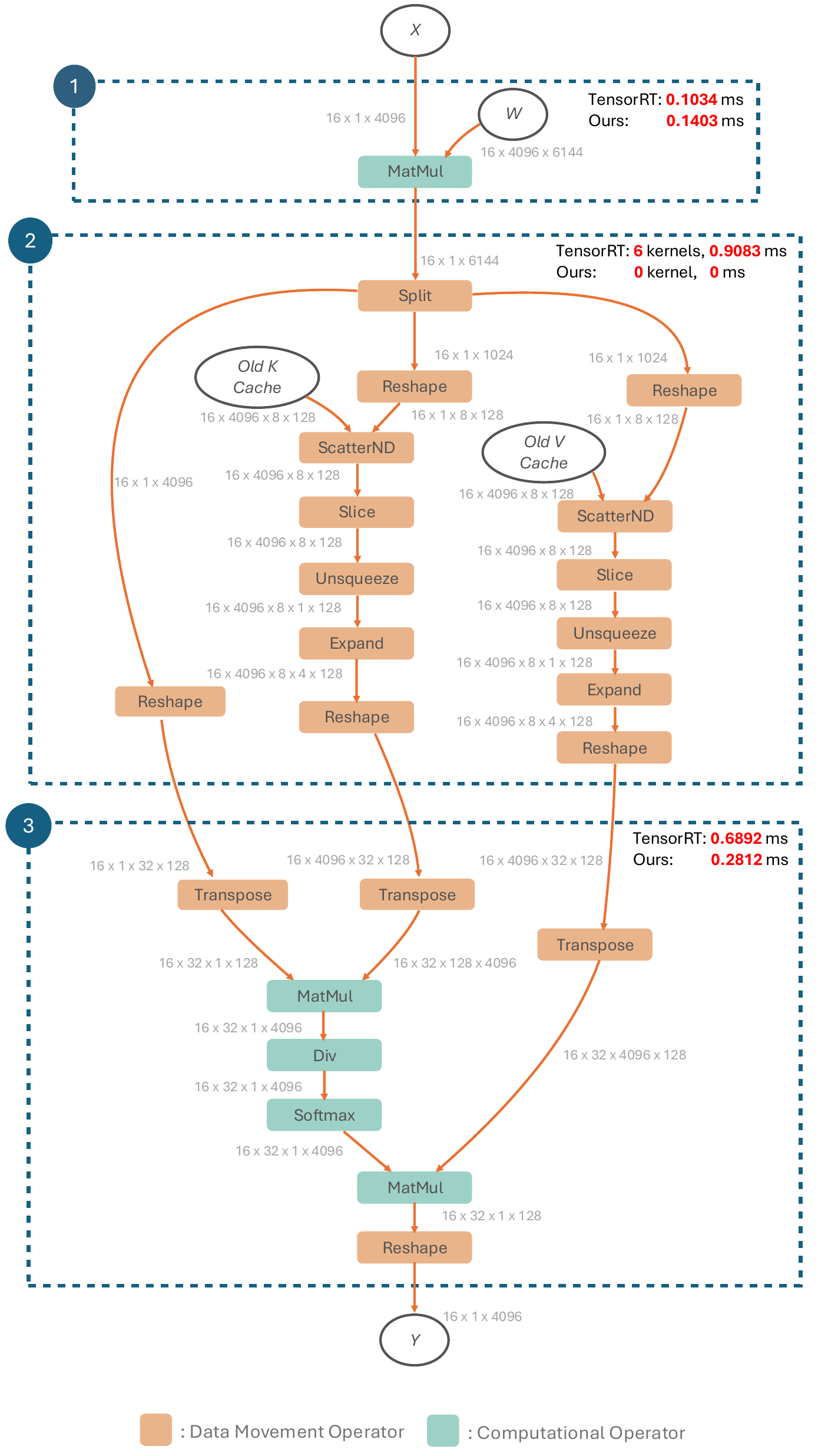}

    \caption{Motivating example: computation graph and latency breakdown for part of a Llama-3 decoder layer. The model size of Llama-3 is 8B, with batch size 16, context length 4096, query length 1 (decoding stage), \texttt{bfloat16} precision. Frame 3 denotes a single kernel fusing all the operators with FlashDecoding. \sys can eliminate all the data movement operators in frame 2 and achieve a $4\times$ speed-up over TensorRT on this subgraph.}
    \label{fig:motivation}
\end{figure}

However, memory technology in modern hardware has not kept pace with these advances in compute capabilities.
\Cref{fig:ratio} illustrates the widening compute-to-memory ratio as newer accelerators are introduced, making memory instructions significantly more expensive than compute instructions.
Moreover, an increasing number of DNN models are becoming memory bound, with performance dominated by memory access requests to and from the accelerator's global memory, which exacerbates the impact of slower memory.
For example, in large language models (LLMs), the incremental decoding stage is memory-bound and can be the bottleneck of end-to-end inference~\cite{sanovar2024lean}.
Memory access requests are determined by the types and the composition of
\emph{data movement operations} in computation graphs.
These operations only transfer data between global memory and the accelerator without performing any computation on tensor data with compute units.
Critically, these inserted data movements between computational operators could introduce substantial extra latency.
For example, \Cref{fig:motivation} demonstrates TensorRT's latency breakdown on a Llama 3 8B~\cite{dubey2024llama} decoder layer.
Between the computation of QKV projection and FlashDecoding~\cite{flashdecoding}, the data movement operators take even more time than other computational operators.
These findings underscore the critical importance of optimizing data movement operations in DNN compilers.

Prior works that propose optimizations for data movement operations in tensor compilers broadly fall under two categories: \emph{layout optimizations} and \emph{operator fusion}, both of which happen in the compiler middle end. These techniques are usually incomplete, targeting only a subset of the data movement operators and missing profitable data movement elimination opportunities that could lead to significant speedups, as detailed below.

\paragraph{Layout optimizations.} ~
The layout of a tensor determines in which order its dimensions are linearized into memory. When a producer operator generates an output tensor with a layout that differs from the expected input layout of the consumer operator, data layout conversion operators need to be inserted, resulting in excessive and avoidable data movement overhead. To find the optimal data layouts that minimize such overheads, previous work first identifies layout-sensitive tensor operators and optimally selects the adjacent layout operators (primarily \texttt{Reshape} and \texttt{Transpose} operators) to balance data movement overhead against computation efficiency~\cite{tensorrt,ivanov2021data,niu2024smartmem}.
However, the tensor operators considered in these works represent only a subset of data movement operations used in DNN models,
overlooking optimization opportunities in other data
movement operators (e.g., \texttt{ScatterND}) that are key for improved performance (see \Cref{sec:motivating-example}).

\paragraph{Operator fusion.}
~During DNN execution, operators usually serve as boundaries of global memory data movement -- all input and output tensors are stored in global memory.
For each operator, its corresponding kernel first loads inputs from global memory into on-chip buffers, performs computations, and then writes results back from on-chip buffers to global memory.
To reduce this costly data movement, existing deep learning frameworks~\cite{tensorrt,chen2018tvm,ansel2024pytorch,abadi2016tensorflow} employ operator fusion, which enables intermediate results to be reused directly from fast on-chip buffers rather than being written to and read from slower global memory.
However, these systems perform fusion using hand-crafted rules that target specific data access patterns (e.g., \cite{dnnfusion}), making this optimization only available for operators that can be fused. As a result, similar to layout optimizations, operator fusion reduces data movement for only a subset of data movement operators, missing important optimization opportunities.

\paragraph{Our Solution.}
~In contrast with these two popular techniques, in this paper, we introduce \sys, a compiler that eliminates unnecessary data movement across all data movement operators.
For example, as shown in \Cref{fig:motivation}, \sys can eliminate all the data movement operators between QKV projection and FlashDecoding.
\sys is comprehensive and works for arbitrary composition of operators in computation graphs, making it generally applicable for all kinds of DNN models. Furthermore, it is complementary to both layout optimization and operator fusion.

The key idea of \sys is to eliminate data movement operators by using \emph{virtual tensors} that store only a mapping function between physical tensors in global memory.
This approach builds on the observation that modern hardware compute units require contiguity only in their local memory buffers, not in global memory.
By relaxing the fully contiguous condition to partially contiguous without compromising memory coalescing, we can significantly eliminate data movement operators.
Virtual tensors enable us to relax the contiguity condition by adding a level of indirection in the accesses to the tensors in global memory.
However, realizing this virtual tensor representation requires addressing a few challenges to fully eliminate unnecessary data movement.

\paragraph{Challenge 1: Kernel creation that facilitates virtual tensors.} ~With the introduction of virtual tensors, the current computational kernels cannot be used in their present form. First, these kernels assume contiguously laid out data in global memory, which is guaranteed by data movement operators. Second, virtual tensors can introduce extra overhead of mapping function and incontiguous memory access if used in current kernel implementations.

We only modify the global memory I/O stages in existing computational kernels with block virtual tensor I/O (\Cref{sec:vt-io}), resulting in minimal changes to the original kernel design and marginal runtime overhead.

\paragraph{Challenge 2: Determining a profitable virtual tensor construction strategy.}
~Creating virtual tensors for all data movement operators is non-trivial.
Different virtual tensor construction strategies lead to different profitabilities.

We solve the problem of finding a profitable option in two parts. First, we introduce a \emph{virtual tensor opportunity graph} to capture all virtual tensor creation options for a given computation graph (\Cref{sec:vtog}). Next, we design a global greedy algorithm that extracts a virtual tensor creation strategy with maximized latency savings (\Cref{sec:greedy}), which works empirically well in the evaluation.

In summary, we make the following contributions.
\begin{itemize}[noitemsep, nolistsep]
    \item We identify a new data movement optimization opportunity by allowing partially contiguous global memory storage in DNN compilation.
    \item We introduce the concept of \emph{virtual tensors} for data movement elimination and efficiently support it in computational kernel generation.
    \item We propose \emph{virtual tensor opportunity graph} and data elimination algorithm to automatically find an efficient strategy for creating virtual tensors.
    \item We implement these concepts in a compilation infrastructure, \sys. Our evaluations show up to $1.93\times$ improvement ($1.28\times$ on average) over existing DNN compilers and up to 60\% memory saving (17.5\% on average) on NVIDIA GPUs.
\end{itemize}

\section{Background}
\label{sec:background}

\noindent We first provide background about tensor stride, data movement operators at the graph-level, their necessity, and implementations of tensor operators at the kernel-level before explaining \sys.

\subsection{Tensor Stride}
\label{subsec:stride}

\noindent To index an $n$-d tensor in the 1D memory, modern tensor compilers need to calculate an integer offset from the index vector $\vec{I}\in\mathbb{N}^n$.
Tensor stride vector $\vec{S}\in\mathbb{N}^n$ indicates the number of elements we need to skip in memory to move from one element to the next in each dimension.
For example, for a 3D tensor of shape $(a,b,c)$, its stride vector $\vec{S}=(bc,c,1)$.
And the indexing will be simply the inner product of the stride vector and the index vector.
In the same example, the offset of $\vec{I}=(i,j,k)$ is $\vec{I}\cdot\vec{S}=i\cdot bc+j\cdot c+k$.

\subsection{Data Movement Operators}

\noindent Data movement operators are a subset of tensor operators responsible for only moving data in global memory.\footnote{Conversely, computational operators are defined by their use of compute units to perform arithmetic operations on the data.}
Examples include operators such as \texttt{Transpose}, \texttt{Split} and \texttt{ScatterND}.\footnote{In this paper, all mentioned operators are defined in ONNX (\url{https://onnx.ai/onnx/operators/}), a widely-used standard DNN format.}
As an example, the semantics of \texttt{Transpose} operator can be given as follows.

\begin{definition}[\texttt{Transpose} Operator]
    \texttt{Transpose} operator takes a single input tensor and generates an output tensor by permuting the dimensions of the input with a given permutation \texttt{perm}: $\forall i\in[0,n),{\rm output\_dimensions}[{\rm perm}[i]]={\rm input\_dimensions}[i]$.
\end{definition}

\Cref{fig:split} also shows a visualized example of \texttt{Split} operator.
It is clear that \texttt{Transpose} and \texttt{Split} do not change values of individual elements, but rather reorganize their positions in the tensor.
This property holds for any data movement operator and we formalize it as follows.

\begin{definition}[Data Movement Operator]
    \label{def:data-movement}
    An operator with input tensors $I_1,\dots,I_n$ and output tensors $O_1,\dots,O_m$ is a data movement operator if and only if there exists a function $F(i,\vec{x})=(j,\vec{y})$ such that $O_i[\vec{x}]=I_j[\vec{y}]$, for all $i\in[1,m],j\in[1,n]$, and $\vec{x},\vec{y}$ are subscripts of $O_i,I_j$, respectively.
\end{definition}

For the \texttt{Transpose} operator, $F(1,\vec{x})=(1,\vec{y})$ where $y_i=perm[x_i]$.
The mapping from outputs to inputs is one-to-one in data movement operators, ensuring that each output tensor element is derived from a unique input tensor element.
Conversely, the mapping from inputs to outputs is one-to-many, indicating that an input tensor element may contribute to multiple output tensor elements.

\subsection{Kernel Implementation of Operators}
\label{subsec:kernel}

\noindent Graph-level tensor operators
are typically implemented as hardware accelerator kernels with all the input and output data physically residing in global memory.
The implementation can be divided into three distinct stages:
\begin{enumerate}[noitemsep, topsep=0pt]
    \item Transfer input data from global memory to on-chip buffers.
    \item Perform computation with compute units, which read inputs from and write outputs to on-chip buffer (not present in data movement operators).
    \item Transfer output from on-chip buffer to global memory.
\end{enumerate}

For example, Listing \ref{fig:triton} shows a matrix multiplication operator implemented in Triton~\cite{tillet2019triton} for GPU execution. In stage 1, two blocks of input matrices (2D-tensors) \texttt{a} and \texttt{b} are loaded as \emph{contiguous} chunks from global memory into on-chip buffers. Stage 2 does the actual compute, in this case, it is an inner product of the loaded matrices. Stage 3 writes back the resultant \texttt{c} matrix from on-chip buffer to global memory also as contiguous chunks.

\begin{figure}
    \lstset{language=Python}
    \begin{lstlisting}
@triton.jit
def matmul_kernel(
    a_ptr, b_ptr, c_ptr, # Pointers to matrices
    ... # Elide some meta-parameters
):
    ... # Elide some pre-processing steps
    accumulator = tl.zeros(BLOCK_SIZE)
    for k in range(0, tl.cdiv(K, BLOCK_SIZE_K)):
        # Stage 1: global -> on-chip
        a = tl.load(a_ptrs, ...)
        b = tl.load(b_ptrs, ...)
        # Stage 2: computation
        accumulator = tl.dot(a, b, accumulator)
        # Advance the ptrs to the next K block
        a_ptrs += BLOCK_SIZE_K * stride_ak
        b_ptrs += BLOCK_SIZE_K * stride_bk
    ... # Elide some post-processing steps
    # Stage 3: on-chip -> global
    tl.store(c_ptrs, c, mask=c_mask)
    \end{lstlisting}
    \captionof{lstlisting}{An example of the three stages: a GPU matrix multiplication kernel written in Triton.}
    \label{fig:triton}
\end{figure}

\subsection{Necessity of Data Movement Operators}

\noindent As seen in \Cref{subsec:kernel}, tensor operators typically assume that the data is \emph{contiguous} in memory during stages 1 and 3 of a kernel implementation. This serves two purposes:

(1) satisfying the requirement of modern compute units for contiguous memory layout of inputs and outputs,\footnote{For example, NVIDIA tensor cores load two matrices of specific shapes from a contiguous memory fragment and then perform fused multiply-add (FMA) operations.}
and (2) optimizing global memory access through memory coalescing and better locality.

Due to the assumption of different kernel implementations, if a producer tensor operator uses a data layout or subsection that differs from its consumer tensor operator, data movement operators must be inserted to reconcile the difference. These data movement operators are necessary to maintain the correctness of the tensor computational graph, but they will introduce extra overhead.

\section{Motivation and Overview}

\subsection{Motivating Example: LLM Decoding}
\label{sec:motivating-example}
\noindent To illustrate the potential for data movement optimization in DNN inference, we examine the decoding stage of Llama 3 8B~\cite{dubey2024llama} as a motivating example.
We profiled a single decoder layer using TensorRT~\cite{tensorrt}, a state-of-the-art DNN compiler on NVIDIA GPUs. \footnote{Similar latency distributions in this section were observed with other DNN compilers such as TorchInductor~\cite{ansel2024pytorch}.}
TensorRT incorporates numerous expert-designed optimization rules and successfully identified the self-attention pattern in the decoder layer, triggering an ad hoc optimization for the entire layer. \footnote{This black-box optimization, called Myelin, fuses the entire transformer decoder layer into a single node and generates highly-optimized custom kernels for this node.\label{footnote:myelin}}

\Cref{fig:motivation} shows the computation graph and TensorRT's latency breakdown on an NVIDIA A100 GPU.
TensorRT first applies graph-level transformations to merge the three query, key, and value (QKV) projections into a single matrix multiplication operation.
It also employs the FlashDecoding algorithm~\cite{flashdecoding} to fuse the computation of self-attention into one kernel.
However, between these two computational kernels, there exists a large number of data movement operators, illustrated in frame 2 of \Cref{fig:motivation}.
The latency of these data movement operators in TensorRT even surpasses the combined latency of the two computational kernels.

We now focus on the purpose of data movement operators and their role in the computation graph.
TensorRT first executes a \texttt{Split} operator to extract the Q, K, V tensors from the result of merged QKV projection.
Next, TensorRT uses \texttt{Reshape} and \texttt{ScatterND} operators to update the KV cache~\cite{pope2023efficiently} with the new K and V tensors.
Subsequently, TensorRT applies the \texttt{Slice} operator to keys and values to obtain a slice containing tokens up to the current sequence length.
As Llama 3 8B employs grouped-query attention~\cite{dubey2024llama}, TensorRT then uses \texttt{Unsqueeze} and \texttt{Expand} operators to align the number of KV heads with the number of query heads.
Finally, TensorRT runs \texttt{Reshape} and \texttt{Transpose} operators to prepare the layout of keys and values for FlashDecoding.
Although operator fusion is available for these data movement operators, TensorRT still requires 6 kernels to complete all the data movements above.

\begin{figure}[t]
    \centering
    \includegraphics[width=\linewidth]{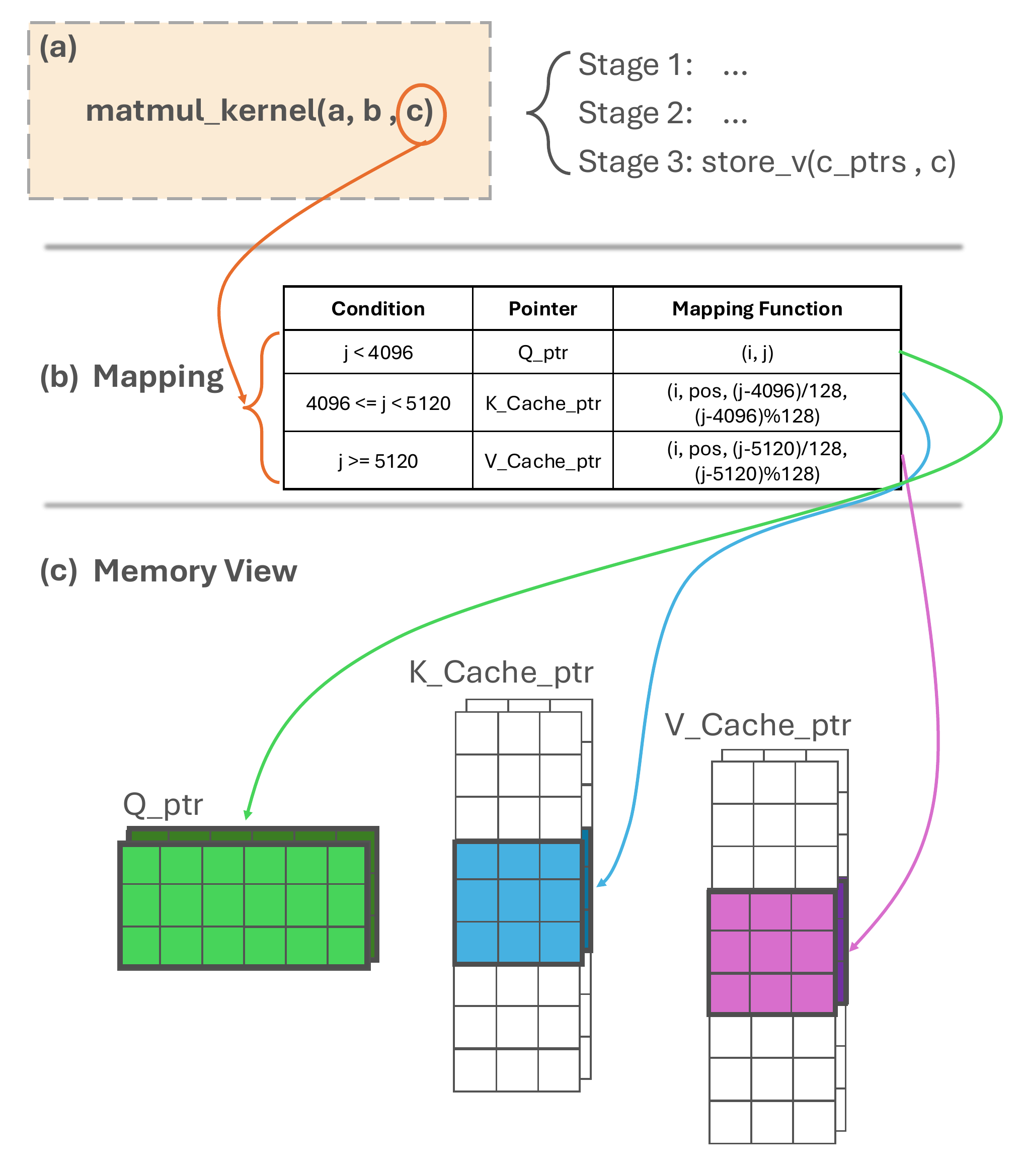}

    \caption{An example of data movement operator elimination with virtual tensors in QKV projection (frame 1 in \Cref{fig:motivation}). The decoding stage is generating the $pos$-th token. Matrix $c[i,j]$ is a virtual tensor of shape (batch size, sum of QKV hidden dimensions), where Q has hidden dimension 4096, and K and V have hidden dimensions 1024 each. For the matrix multiplication kernel (Triton code is in \Cref{fig:triton}), we only need to modify stage 3, and the virtual tensor I/O function \texttt{store\_v} will automatically write the data to proper physical tensor location according to the mapping function.}

    \label{fig:motivation-kernel}
\end{figure}

However, according to the analysis of layout requirements in \Cref{sec:background}, relaxing the fully contiguous condition of global memory storage
may unlock opportunities for data movement operator elimination.
For example, in stage 3 of the \texttt{MatMul} operator used for the merged QKV projection in frame 1, the \texttt{MatMul} kernel can directly write back the results to Q tensor and appropriate locations in KV cache, eliminating the need for \texttt{Split}, \texttt{Reshape} and \texttt{ScatterND} operators, as shown in \Cref{fig:motivation-kernel}.
Since the dimension per attention head (128) is larger than the GPU warp size (32), memory coalescing is not compromised, allowing for both good memory bandwidth in stage 3 and efficient utilization of tensor cores in stage 2.
Similarly, by permitting stage 1 of FlashDecoding to read from not fully contiguous memory, the remaining data movement operators can be eliminated.

Based on these observations, \sys can eliminate \emph{all} the data movement operators between QKV projection and FlashDecoding.
For QKV projection, \sys writes to non-fully contiguous memory, resulting in a slightly slower \texttt{MatMul} kernel compared to TensorRT.
However, \sys does not require any GPU kernel to explicitly execute the data movement operators shown in frame 2 of \Cref{fig:motivation}, significantly reducing data movement overhead.
Furthermore, after \sys's optimization, the FlashDecoding kernel can directly read from the KV cache instead of the duplicated data generated by the \texttt{Expand} operator.
This optimization reduces global memory read overhead by a factor of 4 (the ratio of the \texttt{Expand} operator) and leads to a 2.5$\times$ speed-up of the FlashDecoding kernel.

\subsection{Overview}

\begin{figure}
    \centering
    \includegraphics[width=\linewidth]{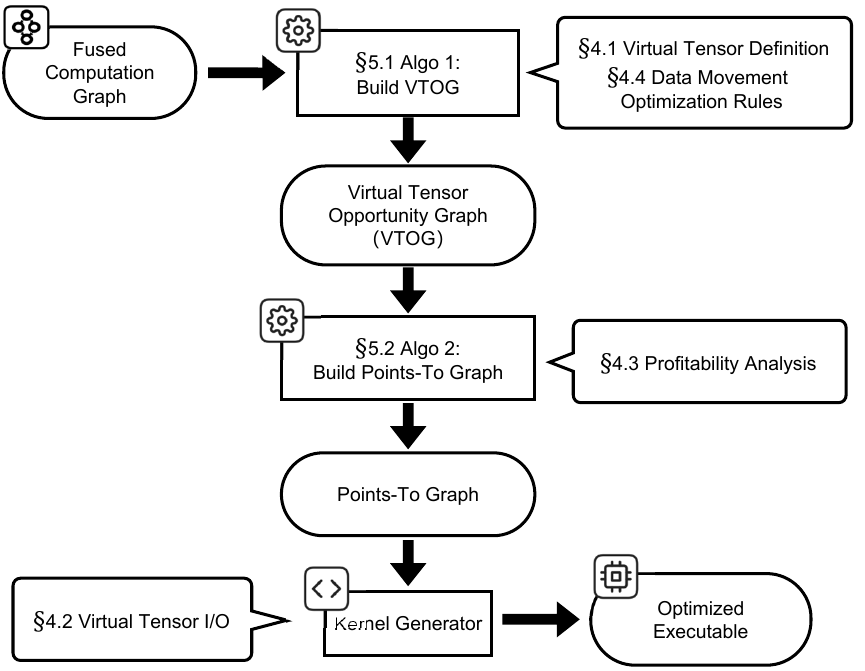}

    \caption{The overview of \sys.}

    \label{fig:overview}
\end{figure}

\noindent \Cref{fig:overview} shows VTC's overview.
The input is a computation graph after operator fusion.
To identify all the virtual tensor creation possibilities, \sys first runs an virtual tensor opportunity graph (VTOG) construction algorithm, based on \sys's virtual tensor definition and data movement optimization rules.
Next, \sys runs a points-to graph construction algorithm to find a profitable virtual tensor strategy, represented by the resulting points-to graph.
Finally, \sys generates an optimized executable according to this selected strategy.

\section{Virtual Tensor}
\noindent \Cref{def:data-movement} reveals that it is possible to eliminate the need for actual data movement by storing only the mapping function $F$, which describes the relationship between the input and output tensors.
By leveraging this insight, we can avoid the overhead associated with explicit data transfers in data movement operators.

In this section, we introduce \sys's virtual tensor notation, which builds upon this observation to efficiently handle data movement operations.
We will show how to handle virtual tensor I/O in \Cref{sec:vt-io} and analyze the profitability of the optimization in \Cref{sec:profitability}.
Finally, we give several examples of how to use virtual tensors to eliminate data movement operators in \Cref{sec:op-example}.

\subsection{Virtual Tensor Definition}
\noindent The key idea behind the virtual tensor technique is to avoid redundant instantiation of tensor data in global memory.
If a tensor is obtained by performing data movement operators on other tensors, we can directly represent it using a mapping function and a set of physical tensor pointers, instead of allocating global memory to store the tensor data.

\begin{definition}[Virtual Tensor]
    A virtual tensor $V$ is a tuple $(F, {P_1, P_2, \dots, P_n})$, where $F$ is a mapping function and ${P_1, P_2, \dots, P_n}$ is a set of physical tensor pointers. The mapping function $F: \vec{x} \rightarrow (j,\vec{y})$ describes the relationship between each virtual tensor index $\vec{x}$ of $V$ and the corresponding physical tensor index $\vec{y}$ of $P_j$, where $j\in [1,n]$.
\end{definition}

\Cref{fig:motivation-kernel} shows an example of virtual tensor in QKV projection.
Virtual tensor $c$ has 3 physical tensor pointers $Q\_ptr$, $K\_Cache\_ptr$, and $V\_Cache\_ptr$, with its mapping function illustrated in subfigure (b).

A virtual tensor provides a contiguous index space that references potentially non-contiguous locations within physical tensors.
Mapping function $F$ links each virtual tensor index to its corresponding physical tensor address.
The virtual tensor technique eliminates data movement operators through two complementary approaches:
(1) creating output virtual tensors for a producer data movement operation that map to input tensors of that operation, or
(2) creating input virtual tensors for a consumer data movement operation that map to output tensors of that operation.
While \Cref{def:data-movement} allows for one-to-many mappings, \sys restricts $F$ to one-to-one mappings. This design choice simplifies implementation and avoids potentially expensive overhead of translating complex one-to-many mapping functions.

\paragraph{Mapping function composition.}
Virtual tensors can be nested: it is possible for a base tensor, which is used to create a virtual tensor, to be a virtual tensor itself.
In such cases, the resulting virtual tensor's mapping function effectively applies the base tensor's mapping followed by the new mapping (e.g., $F_{\rm new}=F_{\rm outer}\circ F_{\rm base}$), representing a chain of transformations.

\subsection{Virtual Tensor I/O}
\label{sec:vt-io}
\noindent \sys's virtual tensor interface seamlessly translates virtual tensor I/O operations into physical global memory I/O operations using the mapping function.
In hardware kernels, global memory I/O is usually performed in a block-wise manner, where multiple threads simultaneously read a block of global memory data.
For most real-world cases, the contiguity of mapping function conditions (e.g., 1024 in \Cref{fig:motivation-kernel}) is larger than and a multiple of block size.
Therefore, each block is likely to read from the same physical tensor, allowing us to directly apply a non-partitioned mapping function to a block of indices, which only introduces marginal overhead.

Virtual tensor provides a high-level abstraction that hides the underlying data movement complexities.
Users can work with virtual tensors as if they were regular tensors, while \sys transparently manages the mapping and access to the physical tensor data behind the scenes.
Since virtual tensors only impact the global memory I/O (stage 1 and 3 in \Cref{fig:triton}) of a hardware accelerator kernels, they leave the computation part (stage 2 in \Cref{fig:triton}) unchanged.
This allows virtual tensors to be effortlessly integrated into existing optimized kernels with minimal modifications.
For example, in \Cref{fig:motivation-kernel}, we can directly pass a virtual tensor mapping to physical Q tensor and KV cache as the output parameter of the projection \texttt{MatMul} kernel.
\sys then transforms the fully contiguous write to this virtual tensor into partially contiguous  writes to the Q tensor and KV cache in \texttt{MatMul} kernel stage 3.

While virtual tensors have the potential to eliminate physical data transfers in data movement operators, the use of mapping functions and non-fully contiguous global memory access patterns still may introduce additional overhead.
In the next subsection, we will delve into the profitability of this optimization technique, analyzing the trade-offs between the benefits of reduced data movement and potential extra costs.

\subsection{Profitability Analysis}
\label{sec:profitability}
\noindent Since the indirect memory access patterns of virtual tensors are fully known at compile time, we can analyze the optimization profitability based on the characteristics of the mapping function.
Similar to the classification of compiler optimizations into Type I (always profitable) and Type II (uncertain profitability)~\cite{mendis2020towards}, we can categorize the profitability of virtual tensor optimizations by analyzing the mapping function.
Without loss of generality, we assume each virtual tensor maps to a single base physical tensor here.

\begin{remark}
    In most DNNs, the final mapping function between the virtual tensor and a physical tensor can be expressed as

    \begin{equation}
        \label{equ:mapping}
        f(\vec{I})=\vec{S}\cdot\vec{I}+b(\vec{I})
    \end{equation}

    where $\vec{I}$ is the index vector, $\vec{S}$ is the stride vector (see \Cref{subsec:stride}) and $b:\mathbb{N}^n\to\mathbb{Z}$ is an index-dependent bias function.
\end{remark}

\begin{remark}
    In ONNX operator set, most data movement operators preserve affine transformations on indexing, which can be achieved by adjusting stride and a \emph{constant} bias. One exception is \texttt{Expand}. For instance, expanding a tensor from shape $(a,b,c)$ to $(a,3b,c)$ and making the expanded tensor virtual yields the offset for index $\vec{I}=(i,j,k)$ as

    \begin{equation}
        i\cdot bc+(j\% b)\cdot c+k=i\cdot bc+j\cdot c+k-\lfloor j/b\rfloor\cdot bc
    \end{equation}

    Here for the mapping function in \Cref{equ:mapping}, $\vec{S}=(bc,c,1)$, $b(\vec{I})=-\lfloor j/b\rfloor\cdot bc$.
\end{remark}

\begin{definition}[Contiguous dimension]
    For an $n$-d virtual tensor with shape $\vec{D}=(D_1,\dots,D_n)$, dimension $d$ is a \emph{contiguous dimension} if its mapping function in \Cref{equ:mapping} satisfies
    \begin{enumerate}
        \item All strides from dimension $d$ onward equal the suffix products of the shape vector: $\forall i\in[d,n]$, $S_i=\prod_{j=i+1}^{n}D_i$
        \item Bias remains constant when the first $d-1$ dimensions are fixed: $\forall \vec{I_1}\in\mathbb{N}^{d-1}$, $\forall\vec{I_2}\in\mathbb{N}^{n-d+1}$, $b([\vec{I_1}\mid\vec{I_2}])\equiv C$.
    \end{enumerate}
\end{definition}

\begin{definition}[Partial contiguity and full contiguity]
    A virtual tensor with minimal contiguous dimension $d$ and shape vector $\vec{D}$ retains \emph{partially contiguous} memory access if $\prod_{i=d}^nD_i$ exceeds the minimal memory transfer unit size (e.g., memory coalescing size in GPUs). A virtual tensor maintains fully contiguous memory access if and only if its minimal contiguous dimension equals 1.
\end{definition}

\begin{theorem}
    \label{thm:type1}
     Optimization is always profitable (Type I) for fully contiguous virtual tensors.
\end{theorem}

\Cref{thm:type1} establishes the conditions for guaranteed profitability.
When the mapping function preserves the original data access pattern with only a constant bias allowed, the virtual tensor optimization can be considered as Type I.
Also, virtual tensors with partial contiguity are likely profitable since they do not break minimal memory transfer granularity and maintain high memory bandwidth.
Otherwise, the profitability is uncertain.
\sys will perform profiling (described in \Cref{sec:greedy}) to decide whether to proceed with the optimization according to actual latency impacts.

\subsection{Virtual Tensor Optimization Rules for Data Movement Operators}
\label{sec:op-example}
\noindent The virtual tensor optimization is designed to eliminate data movement operators.
In this subsection, we present several examples of data movement operators from ONNX, demonstrate how \sys eliminates them using virtual tensors, and discuss the profitability.
All of the data movement operators presented in this subsection are outside the optimization space of previous layout optimizations.

\paragraph{Split}

\begin{figure}
    \centering
    \includegraphics[width=0.7\linewidth]{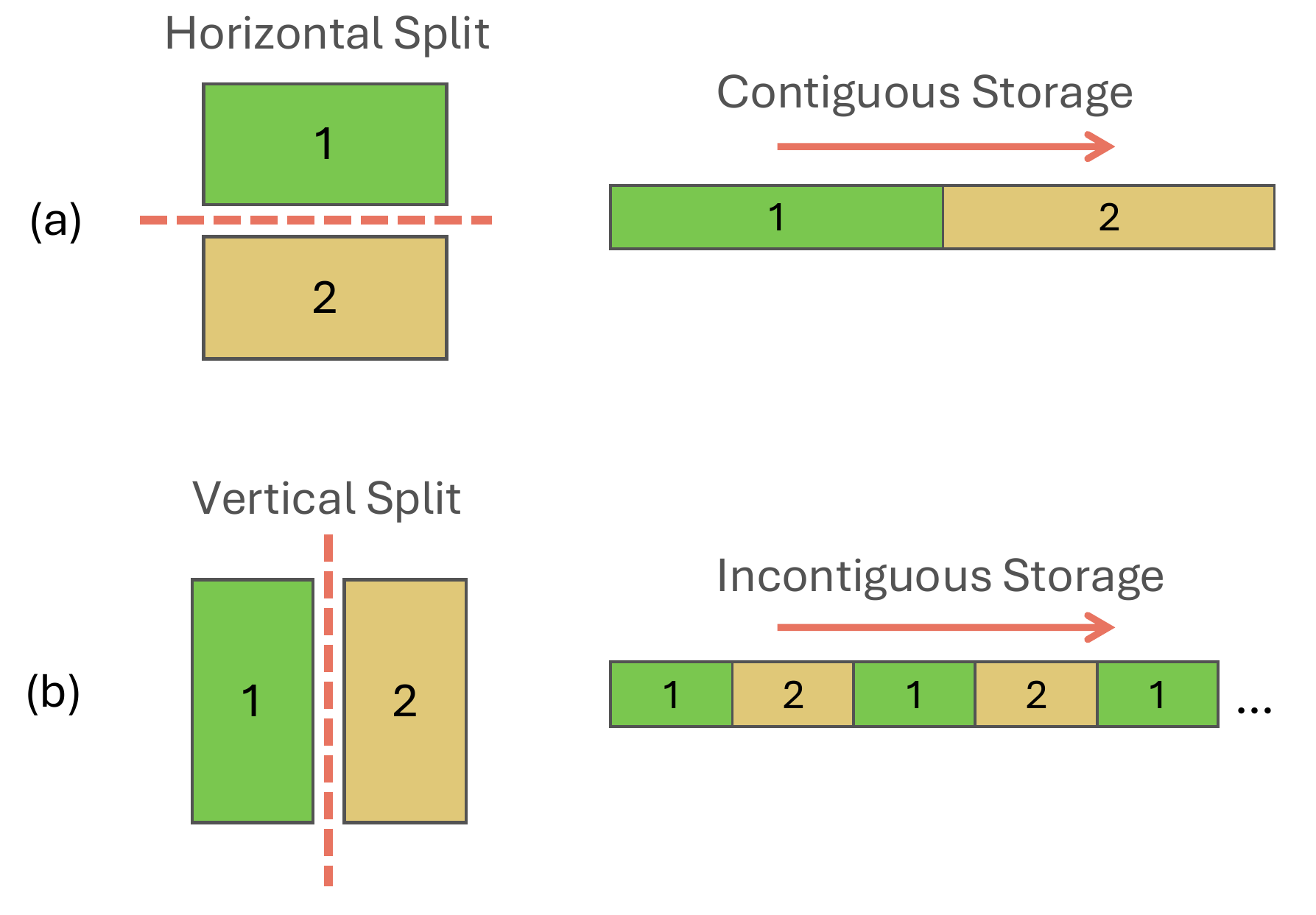}

    \caption{Comparison between contiguous \texttt{Split} (performed on the first axis) and incontiguous \texttt{Split} (performed on other axes) on a 2D matrix stored in row major.}

    \label{fig:split}
\end{figure}

~When optimizing \texttt{Split} using virtual tensors, making input a virtual tensor of outputs is Type II, since the output tensors are stored separately and incontiguously in global memory.
On the other hand, making outputs virtual tensors of input can be either a Type I or Type II optimization, depending on the contiguity of the split.
If the split is performed along the first axis (axis 0), resulting in contiguous output tensors, it is a Type I optimization.
\Cref{fig:split} shows an example of contiguous \texttt{Split} and incontiguous \texttt{Split}.
When the incontiguous split size is large enough, all the two directions of virtual tensor optimizations have partial contiguity.

\paragraph{Expand}
~The \texttt{Expand} operator broadcasts a single input tensor to a specified \texttt{shape}, producing a single output tensor with the same data as the input but with expanded dimensions.
When optimizing \texttt{Expand} using virtual tensors, both directions are Type II.
But if the number of duplicated elements exceeds memory coalescing size, it is almost always profitable.

\paragraph{ScatterND}
~The \texttt{ScatterND} operator takes 3 inputs: \texttt{data}, \texttt{indices}, and \texttt{updates}.
It generates a single output tensor by first initializing it as a clone of the \texttt{data} tensor and then updating the values at the specified indices in \texttt{indices} with the corresponding values from the \texttt{updates} tensor.
There are two ways of virtual tensor optimization:
\begin{enumerate}[noitemsep,topsep=2pt]
    \item Making the output a virtual tensor of \texttt{data}. Since the output tensor is initially a clone of \texttt{data}, it is a Type I optimization.
    \item Making \texttt{updates} a virtual tensor of the output tensor. This optimization is Type II because the \texttt{indices} can be incontiguous. In the LLM motivation example, \texttt{updates} denotes all the locations in KV cache to be updated and is partially contiguous.
\end{enumerate}

We can conduct analysis and define optimization rules similarly for all other data movement operators.
As virtual tensor optimization rules are highly dependent on the operator semantics and the number of data movement operators is limited (e.g., 18 in the ONNX), \sys requires developers to specify the mapping function and possible virtual tensor optimization rules between inputs and outputs for each data movement operator.
Then \sys can automatically analyze profitability and perform virtual tensor optimizations, which will be introduced in the next section.

\section{Automatic Virtual Tensor Construction}
\noindent This section introduces \sys's virtual tensor construction algorithm, which automatically makes intermediate tensors virtual to minimize unnecessary data transfers and improve overall performance.
We first formalize the data movement elimination problem by building a virtual tensor opportunity graph (VTOG), and then introduce the global greedy algorithm to eliminate data movements.

\subsection{Virtual Tensor Opportunity Graph}
\label{sec:vtog}
\noindent \sys's data movement eliminator takes a computation graph after operator fusion as input.

In prior work, all the intermediate tensors, represented by edges in the computation graph, are physically stored in global memory.
The data movement elimination problem aims to find a strategy to make certain intermediate tensors virtual and maximize the end-to-end latency savings.

\sys first builds a virtual tensor opportunity graph (VTOG) to analyze all the virtual tensor possibilities.
\begin{definition}[Virtual Tensor Opportunity Graph]
    \label{def:vtog}
    A virtual tensor opportunity graph (VTOG) is a directed graph $G=(V,E)$. Each node $v\in V$ represents a tensor in the computation graph. Each directed edge $(u,v)\in E$ represents a \emph{direct} virtual tensor possibility, where tensor $u$ can be made a virtual tensor of $v$ by eliminating \emph{only one} data movement operator.
\end{definition}

\begin{figure}
    \centering
    \includegraphics[width=\linewidth]{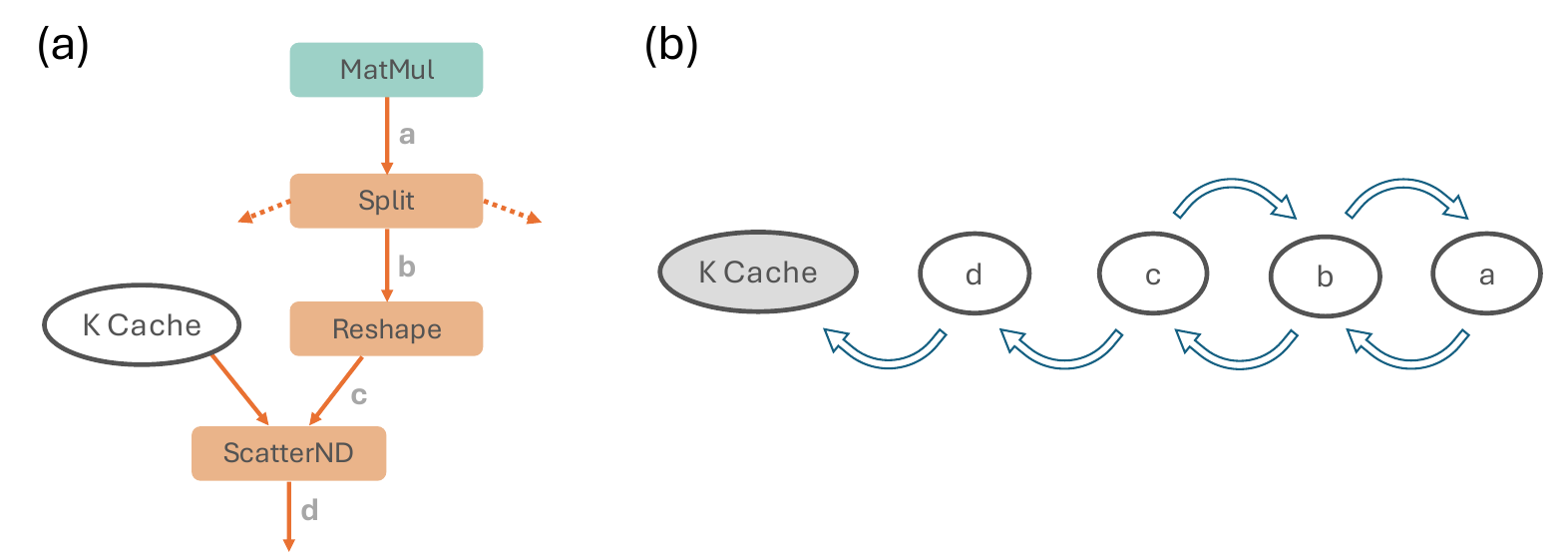}

    \caption{(a) The computation graph. (b) The corresponding virtual tensor opportunity graph. Other outputs of \texttt{Split} operator are ignored.}

    \label{fig:vtog}
\end{figure}

\Cref{fig:vtog} shows an example of a computation graph and its corresponding VTOG.
\texttt{Split} determines the VTOG edges between \texttt{a} and \texttt{b};
\texttt{Reshape} determines the VTOG edges between \texttt{b} and \texttt{c};
and \texttt{ScatterND} determines the VTOG edges between \texttt{c}, \texttt{d} and \texttt{K Cache}.

\begin{figure}
    \centering
    \includegraphics[width=\linewidth]{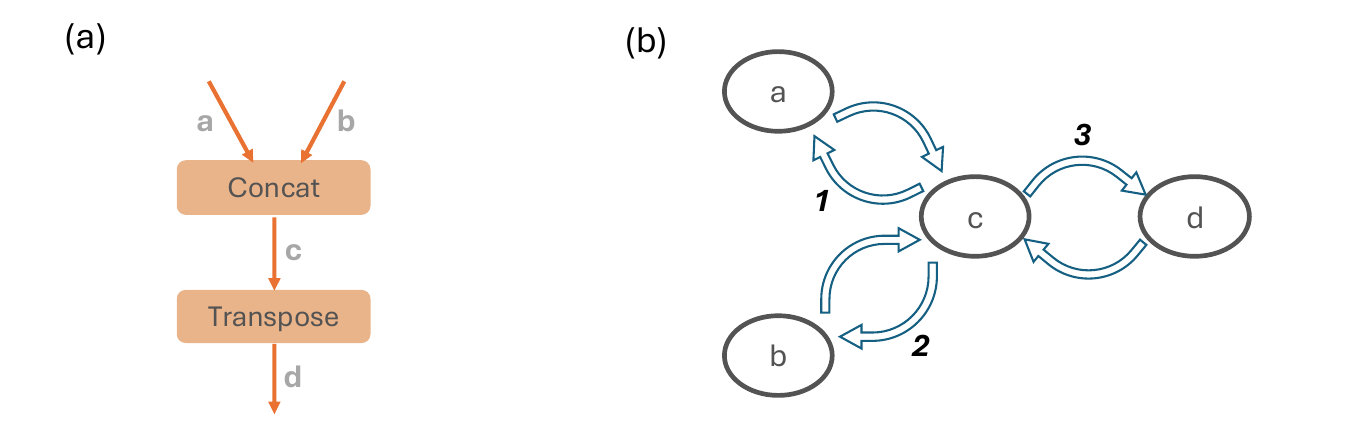}

    \caption{An example of a computation graph and conflicting edges in its VTOG. In (b), edge 1 and edge 3 conflict with each other. Edge 2 and edge 3 also conflict. However, edge 1 and edge 2 can be selected together.}

    \label{fig:conflict}
\end{figure}

For each node in the VTOG, some outgoing edges may conflict with each other, indicating mutually exclusive virtual tensor strategies.
For example, in \Cref{fig:conflict}, edges 1 and 3 conflict because designating \texttt{c} as a virtual tensor of both \texttt{a} and \texttt{d} simultaneously would map some indices of \texttt{c} to multiple physical memory locations, violating the one-to-one property of virtual tensor mapping functions.
In contrast, edges 1 and 2 can coexist when preserving a one-to-one mapping.
We capture all such conflicting edge pairs in a \emph{conflict set}.

\Cref{algo:vtog} shows how to create a VTOG and the conflict sets from a computation graph.
\sys constructs the VTOG by iterating through all data movement operators in the input computation graph, following the predefined virtual tensor rules for each operator (\Cref{sec:op-example}), and creating edges between operator inputs and outputs.

\begin{algorithm}[t]
    \caption{Build VTOG from a computation graph.}
    \label{algo:vtog}
    \small
    \hspace*{\algorithmicindent} \textbf{Input:} A computation graph $G=(V,E)$ \\
    \hspace*{\algorithmicindent} \textbf{Output:} The $\VTOG=(V_\VTOG,E_\VTOG)$ of $G$ and a conflict set $S(v)\ \forall v\in V_\VTOG$
    \begin{algorithmic}[1]
        \State $I=\{i\mid i\text{ is an input tensor of } G\}$
        \State $O=\{o\mid o\text{ is an output tensor of } G\}$
        \State $V_\VTOG=I$
        \For{$v\in V$}
            \State $V_\VTOG=V_\VTOG\cup\{o\mid o\text{ is an output tensor of } v\}$
        \EndFor
        \State $E_\VTOG=\{\}$
        \For{$v\in V$}
            \If{$v$ is a data movement operator}
                \For{$(a,b)\in$\Call{VTRules}{$v$}} \Comment{$a,b$ are input or output tensors of operator $v$ and $a$ can be made a virtual tensor of $b$}
                    \If{$a\notin I$ \textbf{and} $a\notin O$} \Comment{inputs and outputs of $G$ must be physical tensors}
                        \State $E_\VTOG=E_\VTOG\cup\{(a,b)\}$
                    \EndIf
                \EndFor
            \EndIf
        \EndFor
        \For{$v\in V_\VTOG$}
            \State Construct $S(v)$ by enumerating outgoing edges of $v$
        \vspace{0.5em}\EndFor
        \Return $(V_\VTOG,E_\VTOG),\ S$
    \end{algorithmic}
\end{algorithm}

\begin{figure}

    \centering
    \includegraphics[width=\linewidth]{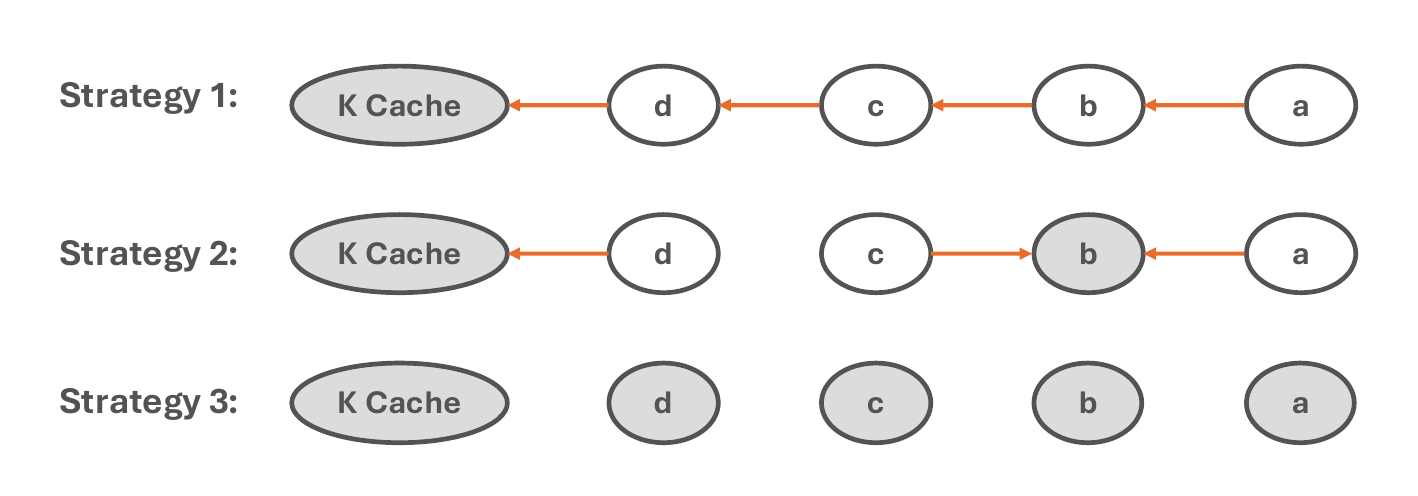}

    \caption{Several valid points-to graph of \Cref{fig:vtog}, representing different virtual tensor strategies. Physical tensors are marked gray in each strategy.}

    \label{fig:ptg}
\end{figure}

When all outgoing edges of each node in a VTOG are mutually compatible, the VTOG becomes a \emph{points-to graph}, similar to that used in traditional compiler pointer analysis~\cite{andersen1994program}.
The points-to graph has one-to-one correspondence with a virtual tensor strategy for the entire computation graph.
For example, \Cref{fig:ptg} shows several possible points-to graphs derived from the VTOG in \Cref{fig:vtog}.
Strategy 3 has no edges, indicating that all the tensors are physical.
Strategy 2 has \texttt{b} and \texttt{K cache} as physical tensors, making \texttt{a} and \texttt{c} virtual tensors of \texttt{b}, and \texttt{d} a virtual tensor of \texttt{K cache}.
Strategy 1 makes \texttt{a}, \texttt{b}, \texttt{c}, and \texttt{d} all virtual tensors of \texttt{K cache}.
Note that \texttt{K cache} is an input tensor of the computation graph, so it must reside in physical memory.

Using an inductive argument, it is not hard to show the computational equivalence of the computation graph and the virtualization strategy corresponding to a points-to graph derived from the VTOG.
As evident from the definition, a points-to graph can be derived from a VTOG by removing conflicting edges.
However, different points-to graphs can lead to different end-to-end latency gains.
The following subsection introduces \sys's points-to graph construction algorithm, which aims to obtain a points-to graph representing an efficient virtual tensor strategy.

\subsection{Global Greedy Algorithm}
\label{sec:greedy}

\noindent The objective of our data movement elimination algorithm is to maximize the latency saving for each virtual tensor strategy, which can be formalized as follows.
Let $(E_\VTOG, \mathbb{F}, \ell)$ denote the virtual tensor search space corresponding to a VTOG.
Here each edge in $E_\VTOG$ denotes a single virtualization opportunity, $\mathbb{F}$ denotes a set of all points-to graphs and $\ell: \mathbb{F} \rightarrow \mathbb{R}$ is the latency saving function that maps every points-to graph $C \in \mathbb{F}$ to the latency saving corresponding to virtual tensor strategy realized by $C$.
Our goal is to find the optimal points-to graph
{\setlength{\abovedisplayskip}{3pt}
\setlength{\belowdisplayskip}{3pt}
\begin{equation}
\label{e:latencyfunction}
C^* = \text{arg}\,\max_{\mathbb{F}} \left\{ \ell(C) \right\}
\end{equation}}
\par\noindent\Cref{e:latencyfunction} is NP-hard as $\ell$ is a complex set function without any structure, which can only be measured through profiling on real hardware.
In some situations we can expect that $\ell$ is a submodular function; for example, when all our operations are Type I (see \Cref{sec:profitability}).
Unfortunately, it is not true for more general VTOGs.

Despite this, we take inspiration from the submodular function maximization literature, and
design a global greedy algorithm to approximately solve our problem.
It is well-known that the global greedy algorithm, which at each iteration adds the element that maximizes the discrete derivative of a submodular function $\ell$, achieves a constant approximation to the optimal solution for a broad range of settings; we refer readers to \cite{krause2014submodular} for an introductory survey.

\Cref{algo:greedy} shows \sys's points-to graph construction procedure.
With all tensors stored in physical memory initially, it iteratively creates new virtual tensors.
Specifically, the algorithm maintains an anchor node set $A$ and a currently selected edge set $C$.
In each iteration, it adds a new node $v_c$ into the anchor node set by adding some non-conflicting edges between $v_c$ and $A$ into the currently selected edge set.
For each edge $e\in E_\VTOG$, we define its \emph{discrete derivative with respect to $\ell$} as
{\setlength{\abovedisplayskip}{3pt}
\setlength{\belowdisplayskip}{3pt}
\begin{equation}
    w(e):=\ell(C\cup\{e\})-\ell(C),
\end{equation}
}
\par\noindent which denotes the latency saving of merely adding edge $e$ to $C$.
The greedy strategy picks the node $v_c$ that maximizes the sum of latency savings across its incident non-conflicting edges in the current iteration, then updates the sets $A$ and $C$.
Subsequently, since creating a new virtual tensor updates $C$, the algorithm recalculates $w(e)$ for each edge $e$ pointing to $v_c$ by profiling.
If the maximum latency saving is negative or all the nodes are added to $A$, the algorithm terminates.

\begin{algorithm}[t]

    \caption{An iterative global greedy algorithm to build a points-to graph from a VTOG.}
    \label{algo:greedy}
    \small
    \hspace*{\algorithmicindent} \textbf{Input:} $\VTOG=(V_\VTOG,E_\VTOG)$ and conflict set $S(v)$ \\
    \hspace*{\algorithmicindent} \textbf{Output:} The points-to graph $\PTG=(V_\PTG,E_\PTG)$ of $\VTOG$ with maximized latency savings
    \begin{algorithmic}[1]
    \setlength{\itemsep}{0.0em}
        \State $A=\{\}$, $C=\{\}$
        \For{$v\in V_\VTOG$}
            \If{$v$ does not have outgoing edges}
                \State $A=A\cup \{v\}$
            \EndIf
        \EndFor
        \For{$e\in E_\VTOG$}
            \State Profile and get $w(e)$ \label{line:profile1}
        \EndFor
        \State $\text{tot\_saving}=0$
        \While{$A\ne V_\VTOG$}
            \State $\text{max\_saving}=-\infty$
            \For{$v\in V_\VTOG\setminus A$}
                \State $P,s=$\Call{MaxEdges}{$v,C(v),A$} \Comment{$P$ is a set of non-conflicting edges between $v$ and $A$ with maximum sum of marginal latency savings $s$}
                \If{$s>\text{max\_saving}$}
                    \State $\text{max\_saving}=s$, $P_c=P$, $v_c=v$
                \EndIf
            \EndFor
            \If{$\text{max\_saving}<0$}
                \State\textbf{break}
            \EndIf
            \State $A=A\cup\{v_c\}$, $C=C\cup P_c$
            \State $\text{tot\_saving}=\text{tot\_saving}+\text{max\_saving}$
            \For{$e=(t,v_c)\in E_\VTOG$}
                \State Profile and update $w(e)$ \label{line:profile2}
            \EndFor
        \vspace{0.5em}\EndWhile
        \Return $(V_\VTOG,C)$, tot\_saving
    \end{algorithmic}
\end{algorithm}

\paragraph{Complexity analysis.}
~\Cref{algo:greedy} executes a maximum of $O(|V_\VTOG|)$ iterations.
In each iteration, it performs at most $O(|E_\VTOG|+|V_\VTOG|)$ calculations.
Based on the VTOG definition \ref{def:vtog} and its construction algorithm \ref{algo:vtog}, $|E_\VTOG|$ is linear with $|V|$ and $|V_\VTOG|<|E|$.
In typical DNN computation graphs, $O(|V|)=O(|E|)$, so $O(|V_\VTOG|)=O(|E_\VTOG|)=O(|V|)$.
Consequently, the complexity of \sys's data movement elimination algorithm is the product of the maximum number of iterations and the cost per iteration, i.e. $O\left(|V|^2\right)$.

\section{Implementation}
\label{sec:impl}

\noindent
We implemented \sys on top of Triton~\cite{tillet2019triton} and TorchInductor~\cite{ansel2024pytorch}.
We first added virtual tensor support in Triton by overloading Triton's global memory I/O functions, \texttt{tl.load} and \texttt{tl.store}, enabling them to handle virtual tensor I/O operations.
Mapping function is represented in the form of \Cref{equ:mapping}.
Our overloaded operators consume base physical tensor pointers along with mapping functions that determine how virtual tensors map to physical memory.
Since all indirect memory access patterns through virtual tensors are fully known at compile time, we implemented overloaded \texttt{tl.load} and \texttt{tl.store} by generating Python code specialized for each virtual tensor I/O, avoiding most of the indirection overheads.

We integrated \sys's data movement eliminator into TorchInductor through two passes: an analysis pass and a transformation pass.
The analysis pass automatically generates a virtual tensor strategy with \Cref{algo:vtog,algo:greedy}.
This entails identifying which physical tensors to promote to virtual tensors as well as computing mapping functions of promoted virtual tensors to their physical indices.
The transformation pass then operates in two stages:
(1) It mutates TorchInductor's IR nodes with virtual tensor awareness and removes any unnecessary data-movement operators.
We modified TorchInductor's lowering strategy to correctly lower our overloaded \texttt{ops.load} and \texttt{ops.store} operators to Triton.
(2) Once lowered to Triton, we generate the specialized Python code using the method mentioned earlier.
The resulting Triton code is then compiled to produce an optimized hardware executable.

\sys's automatic virtual tensor construction employs a polynomial-time greedy algorithm rather than exhaustively searching in an exponential solution space.
During compilation, profiling (lines \ref{line:profile1} and \ref{line:profile2} in \Cref{algo:greedy}) dominates the time cost.
Since \sys only modifies data movements in stages 1 and 3, no autotuning is needed for each kernel.
The profiling overhead is less than 10 seconds per configuration, and the end-to-end compilation requires only a polynomial number of profiling runs, resulting in total compilation time under 10 minutes for all the models we have tested, which is on par with the compilation time of ML compilers~\cite{shi2023welder,zhao2024felix,zhang2023cocktailer}.

As \sys focuses solely on optimizing data movement and does not alter the computation logic, it maintains end-to-end \emph{numeric equivalence} with the original compiler, ensuring zero precision loss.

\section{Evaluation}
\label{sec:eval}

\subsection{Experimental Setup}
\paragraph{Platform.}
~We conduct our evaluation on an A100 server and an H100 server.
The A100 server is equipped with an Intel Xeon Platinum 8358 CPU and an NVIDIA A100 80GB PCIe GPU.
The H100 server is equipped with an AMD EPYC 9454 Processor and an NVIDIA H100 NVL GPU.
Both servers run Ubuntu 22.04, NVIDIA driver version 535.183.06 and CUDA version 12.1.

\paragraph{Workloads.}
~We use five real-world DNNs from various domains to evaluate \sys.
Llama 3~\cite{dubey2024llama} and Gemma 2~\cite{team2024gemma} are transformer-based LLMs with grouped-query attention and local-global attention, respectively.
EfficientViT~\cite{cai2023efficientvit} is a vision transformer backbone for high-resolution dense prediction with linear attention.
YOLOv11~\cite{khanam2024yolov11} is a powerful and efficient convolutional neural network (CNN) designed for a broad range of vision applications.
ShuffleNet~\cite{zhang2018shufflenet} is a CNN optimized for mobile devices.
\Cref{tab:workloads} provides detailed configurations for each model.
We use PyTorch implementation for all the models.
Except ShuffleNet, all the implementations are from official repositories.
Following the common practice in previous ML compiler research~\cite{zheng2020ansor,wang2021pet,zheng2023einnet}, we evaluate all models with batch size 1 and 16.

\begin{table}
\caption{Configurations of evaluated models.}
\label{tab:workloads}
\resizebox{\linewidth}{!}{
\centering
\begin{tabular}{c|ccc}
\toprule
Model & Component & Input Specification & Precision \\
\midrule
Llama 3 8B & \multirow{2}*{Decoder Layer} & Query Length: 1, & \multirow{2}*{BF16} \\
Gemma 2 9B & & Context Length: 4096 & \\
\midrule
EfficientViT-Base & Attention Block & Resolution: 4096 & \multirow{3}*{TF32} \\
YOLOv11n & C3K2 Block & Resolution: 640 & \\
ShuffleNet & ShuffleUnit & Resolution: 224 & \\
\bottomrule
\end{tabular}
}

\end{table}

\subsection{End-to-end Performance}
\begin{figure*}[h!]

        \centering
        \includegraphics[width=\linewidth]{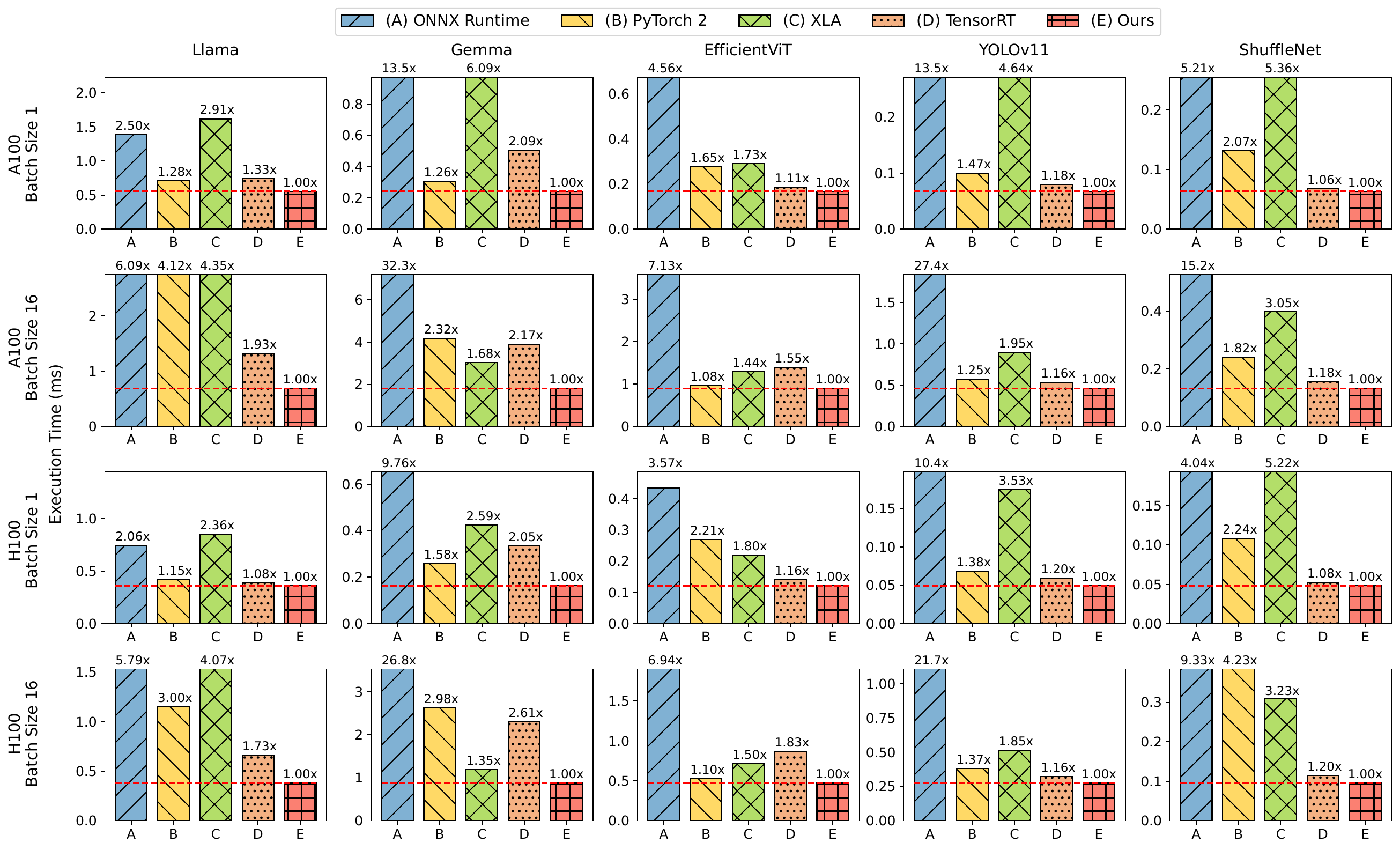}

        \caption{End-to-end inference latency comparison on a single NVIDIA A100 GPU and H100 GPU with batch sizes 1 and 16. Bars over $4\times$ of \sys's latency are truncated.}

        \label{fig:e2e}

\end{figure*}

\noindent We first evaluate the end-to-end inference latency on a single A100 or H100 GPU and compare \sys with PyTorch 2.6.0~\cite{ansel2024pytorch} (with \texttt{torch.compile} enabled), ONNX Runtime 1.21.1~\cite{onnxruntime}, XLA\footnote{Since our benchmark workloads are implemented in PyTorch, we use \texttt{torch\_xla} 2.6.0 package to run PyTorch models with XLA backend.}~\cite{xla} and TensorRT 10~\cite{tensorrt}.
\Cref{fig:e2e} presents the end-to-end inference latency results with different batch sizes on A100 and H100, respectively.
\sys achieves up to $1.93\times$ speed-up ($1.28\times$ on average) over the best of the four baselines (mostly TensorRT or PyTorch) in each benchmark.

During the evaluation, TensorRT successfully identifies the attention pattern for all Transformer-based models (Llama, Gemma and EfficientViT) and triggered highly optimized implementations for the entire model. \footref{footnote:myelin}
However, \sys still demonstrates an average speed-up of $1.36\times$.
This is notably higher than the average speed-up of $1.15\times$ observed for the CNN models (YOLOv11 and ShuffleNet), indicating that \sys effectively capitalizes on optimization opportunities inherent in newer model architectures like Transformers.

\sys achieves even greater speedup on H100 GPUs as compared to A100 GPUs for 7 cases out of the 10 benchmark cases (covering 5 models at 2 different batch sizes).
This trend suggests that the optimization techniques employed by \sys are particularly effective on newer hardware accelerators, where the gap between compute power and memory bandwidth is getting larger (see \Cref{fig:ratio}).

\subsection{Memory Footprint}
\noindent We also measure the inference GPU memory footprint after \sys's optimization compared to the PyTorch baseline (as \sys is built upon TorchInductor) using the PyTorch's built-in memory monitor.
As shown in \Cref{tab:memory}, even when optimizing for maximized latency reduction, \sys still achieves up to 60\% peak memory saving (17.5\% on average).
This demonstrates that virtual tensor optimization inherently saves memory -- it avoids storing full physical tensor data by representing virtual tensors with tensor pointers and mapping functions, and \sys handles virtual tensor I/O through code generation with no extra memory overhead.
Analyzing the results in detail, we find that \sys successfully makes many large intermediate tensors virtual (e.g., the output of \texttt{ScatterND} in \Cref{fig:motivation}), thereby reducing peak memory usage.
Moreover, this highlights \sys's potential for more aggressive memory saving with relaxed latency constraints by reconfiguring the function $\ell$ in \Cref{sec:greedy} to measure memory reduction, offering greater benefits for model training and deployment on memory-constrained hardware.

\begin{table}
\caption{Peak GPU memory consumption of PyTorch and \sys during end-to-end inference with batch size 1 and 16 on A100. All numbers are in Megabytes.}

\label{tab:memory}
\resizebox{\linewidth}{!}{
\centering
\begin{tabular}{c|cccccc}
\toprule
Model & Llama & Gemma & EfficientViT & YOLOv11 & ShuffleNet \\
\midrule
PyTorch (BS=1) & 528.8 & 480.7 & 59.0 & 32.8 & 10.6 \\
Ours (BS=1) & \textbf{461.8} & \textbf{436.1} & \textbf{55.6} & \textbf{26.7} & \textbf{10.4} \\
Savings (\%) & \textbf{12.7\%} & \textbf{9.3\%} & \textbf{5.7\%} & \textbf{18.5\%} & \textbf{0.9\%} \\
\midrule
PyTorch (BS=16) & 1787.3 & 1491.6 & 798.1 & 529.0 & 41.0 \\
Ours (BS=16) & \textbf{714.5} & \textbf{776.4} & \textbf{773.9} & \textbf{450.9} & \textbf{39.9} \\
Savings (\%) & \textbf{60.0\%} & \textbf{48.0\%} & \textbf{3.0\%} & \textbf{14.8\%} & \textbf{2.6\%} \\
\bottomrule
\end{tabular}
}

\end{table}

\subsection{Latency Breakdown Analysis}
\label{sec:breakdown}
\begin{figure*}
    \begin{minipage}[t]{0.6\textwidth}
    \centering
    \includegraphics[width=\linewidth]{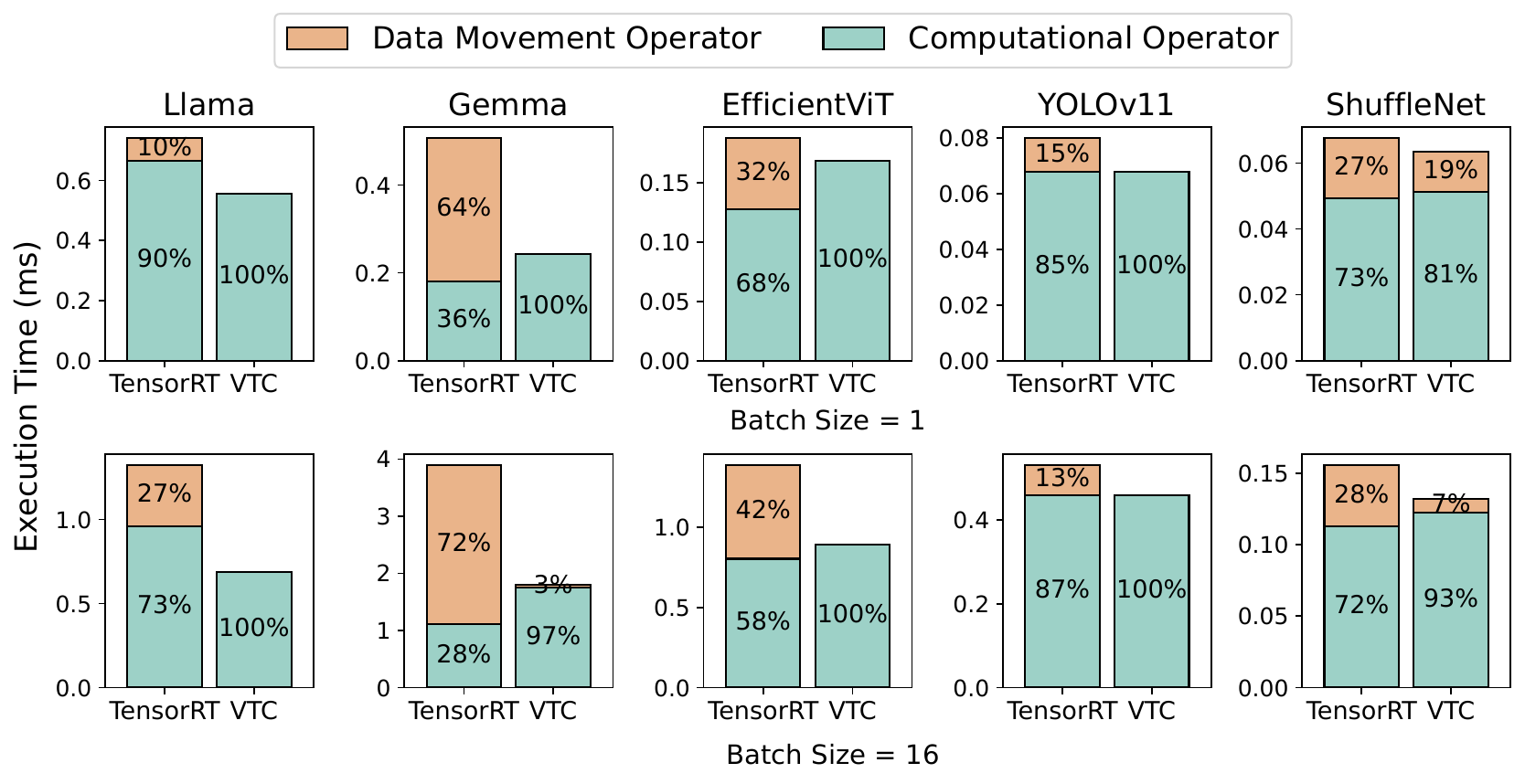}

    \caption{Breakdown of latency proportions for data movement and computational operators in \sys compared with TensorRT.}
    \label{fig:breakdown}
    \end{minipage}
    \hfill
    \begin{minipage}[t]{0.35\textwidth}
        \centering
        \includegraphics[width=0.6\linewidth]{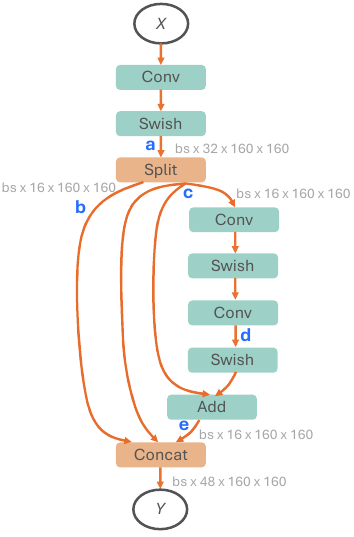}

        \caption{Computation graph of YOLOv11 C3K2 Block. \sys eliminates both data movement operators \texttt{Split} and \texttt{Concat} by making \texttt{a}, \texttt{b}, \texttt{c} and \texttt{e} virtual tensors of \texttt{Y}.}

        \label{fig:yolo}
    \end{minipage}
\end{figure*}

\noindent Since TensorRT is the fastest baseline in most cases, we compare the latency breakdown of data movement operators and computational operators for \sys and TensorRT on A100.
The results in \Cref{fig:breakdown} reveal a substantial reduction in the latency proportion from data movement after \sys's optimization.
In 7 out of 10 evaluated cases, \sys eliminates data movement operators entirely.
Moreover, the analysis highlights that data movement operators account for a larger proportion of the overall latency in Transformer-based models compared to CNNs, enabling \sys to get better speed-ups.

However, in most of the benchmarks, \sys's latency on computational operators is higher than TensorRT's.
There are two possible reasons:
(1) \sys's computational kernels may perform virtual tensor I/O in non-contiguous physical addresses, introducing extra overhead in computational kernels (e.g., the \texttt{MatMul} kernel in frame 1 of \Cref{fig:motivation}).
(2) \sys is built upon Triton and TorchInductor, while TensorRT has more highly optimized implementations of computational kernels.
This indicates that the performance gains from eliminating data movement significantly outweigh the slowdown incurred in computational operators.

\subsection{Comparison with vLLM}
\noindent As \Cref{fig:motivation} shows, \sys eliminates all data movement operators between QKV projection and attention in LLM decoder layer.
Compared with existing compilers such as TensorRT and TorchInductor, the subsequent computational kernel FlashDecoding can launch directly after the previous computational kernel MatMul, improving the utilization rate of compute units without stalling on data movement operators.

To demonstrate the effectiveness of this optimization over a stronger LLM inference baseline beyond compilers, we compare \sys with a state-of-the-art LLM serving framework vLLM V1~\cite{kwon2023efficient} on Llama 3 8B inference.
Since vLLM is PyTorch-based, we can seamlessly integrate \sys into it.
We use similar settings (input sequence length 4096 and BF16 precision)\footnote{vLLM V1 removes traditional notions of batch size and stage separation between prefill and decode through token-based iterative decoding. For our experiments, we construct an input batch of 4096 tokens and measure the corresponding attention latency.} for evaluation on A100 and H100 GPUs.

\Cref{tab:vllm-results} shows that vLLM has \emph{manually} optimized many data movement operators from frame 2 of \Cref{fig:motivation} -- the \textit{KV Cache Update} row in \Cref{tab:vllm-results} contributes to a small fraction of latency compared to compiler baselines.
This poses a great challenge for \sys to compete, as vLLM is heavily optimized by domain experts on \emph{specific} LLMs, while \sys is an \emph{automated} compiler for \emph{general} ML models.
Nevertheless, \sys still identifies a redundant data movement involving the temporary tensor that stores QKV projection results, detailed in \Cref{tab:vllm-explain}.
This optimization yields a 1.043x speed-up on the \Cref{fig:motivation} computation graph and 1.011x end-to-end speed-up over vLLM on A100.
However, no speed-up is observed on H100 since vLLM's cuBLAS MatMul kernel (QKV projection) significantly outperforms \sys's Triton kernel, and \sys's profile-guided virtual tensor construction automatically skips negative optimizations.
Forcing the optimization from \Cref{tab:vllm-explain} on H100 causes an 8\% performance degradation.
We believe this gap could be bridged by integrating \sys into more optimized hardware-specific kernel libraries like CUTLASS in the future.

\begin{table}[t!]
\caption{Comparison of Llama inference latency (ms) between \sys and vLLM. \textit{QKV Projection}, \textit{KV Cache Update} and \textit{Attention} correspond to the 3 frames in \Cref{fig:motivation} in the same order. \textit{Total} shows the latency on the \Cref{fig:motivation} subgraph. \textit{End-to-end} shows latency for the entire decoder layer. \textit{Speed-up} uses vLLM as the baseline. On H100, \sys defaults to identical behavior as vLLM, while enforcing optimization in \Cref{tab:vllm-explain} degrades performance.}

\label{tab:vllm-results}
\resizebox{\linewidth}{!}{
\begin{tabular}{c|cc|ccc}
\toprule
& \multicolumn{2}{c|}{A100} & \multicolumn{3}{c}{H100} \\
& vLLM & \sys & vLLM & VTC & \makecell{\sys w/\\enforced opt.} \\ \midrule
QKV Projection & 0.959 & \textbf{0.914} & 0.401 & 0.401 & 0.714 \\
KV Cache Update & 0.037 & \textbf{0} & 0.027 & 0.027 & \textbf{0} \\
Attention & 0.976 & 0.976 & 0.438 & 0.438 & 0.438 \\
Total & 1.972 & \textbf{1.890} & 0.866 & 0.866 & 1.152 \\
Speed-up & 1.000x & \textbf{1.043x} & 1.000x & 1.000x & 0.752x \\ \midrule
End-to-end & 9.152 & \textbf{9.055} & 4.230 & 4.230 & 4.600 \\
Speed-up & 1.000x & \textbf{1.011x} & 1.000x & 1.000x & 0.920x \\ \bottomrule
\end{tabular}
}

\end{table}

\begin{table}[t!]
\vspace{-3em}
\caption{Data movement optimization in \sys compared to vLLM. As defined in \Cref{subsec:kernel}, \textit{Stage 1 Input} and \textit{Stage 3 Output} represent tensors stored in global memory that kernels read from and write to, respectively. vLLM allocates a temporary tensor to hold merged QKV projection values and splits it to update KV cache in an additional data movement kernel. \sys eliminates this overhead by making the temporary tensor virtual, directly writing results to Q tensor and KV cache.}

\label{tab:vllm-explain}
\resizebox{\linewidth}{!}{
\begin{tabular}{c|ccc}
\toprule
vLLM & Stage 1 Input & Stage 2 & Stage 3 Output \\ \midrule
QKV Projection & Proj Input & MatMul & \textbf{Temporary Tensor} \\
KV Cache Update & \textbf{Temporary Tensor} & N/A & \textbf{Q, KV Cache} \\
Attention & Q, KV Cache & FlashAttn & Attn Output \\ \bottomrule \toprule
\sys & Stage 1 Input & Stage 2 & Stage 3 Output \\ \midrule
QKV Projection & Proj Input & MatMul & \textbf{Q, KV Cache} \\
Attention & Q, KV Cache & FlashAttn & Attn Output \\
\bottomrule
\end{tabular}
}
\end{table}

\subsection{Case Studies}
\label{sec:case}
\noindent To gain a deeper understanding of how \sys eliminates data movement operators, we study some optimization cases found by \sys in detail.

\begin{figure}[h]

    \centering
    \includegraphics[width=\linewidth]{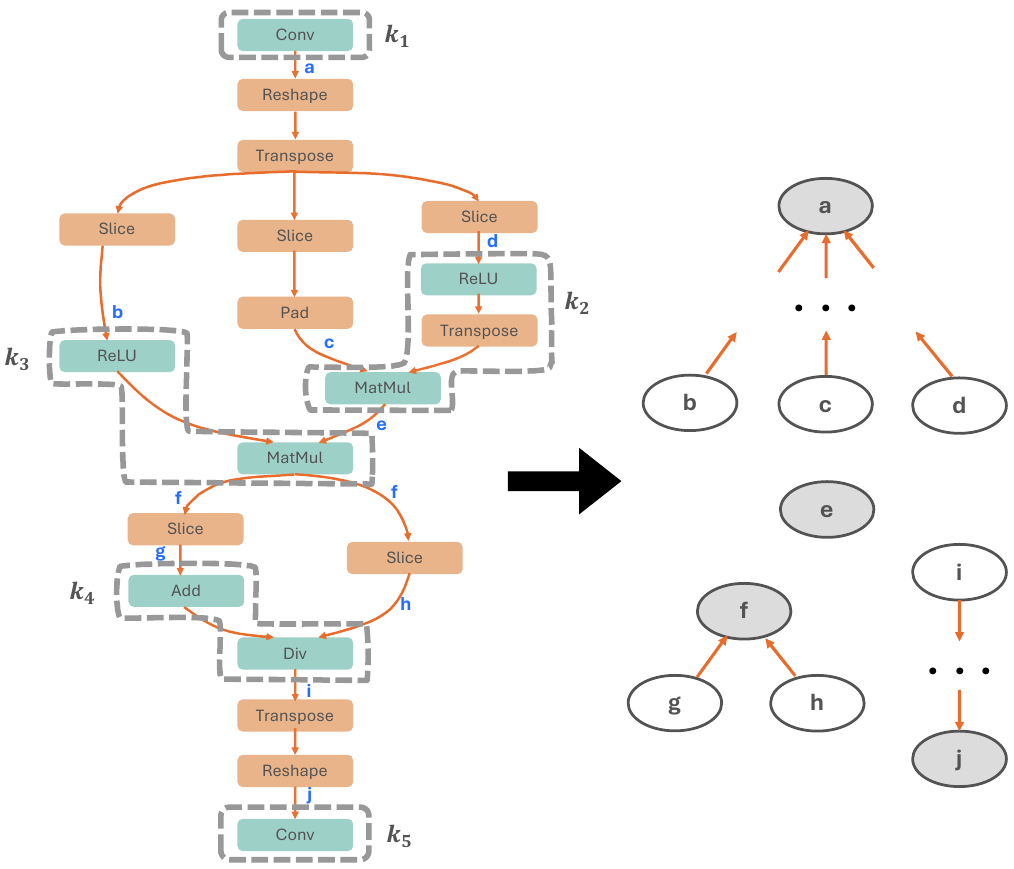}

    \caption{\sys generates 5 kernels ($k_1$--$k_5$) for EfficientViT attention block with batch size 16. All the unframed data movement operators are eliminated with virtual tensor. The figure on the right shows the virtual tensor strategy represented by a points-to graph.}
    \label{fig:efficientvit}

\end{figure}

\paragraph{EfficientViT.}
~\Cref{fig:efficientvit} illustrates \sys's generated kernels and the points-to graph for EfficientViT attention block with batch size 16.
All 5 kernels are computational kernels, and \sys eliminates data movement operators between these kernels by employing a virtual tensor strategy shown in the points-to graph.
Among the tensors, only \texttt{a}, \texttt{e}, \texttt{f}, and \texttt{j} are physically stored.
$k_2$ and $k_3$ directly read their inputs from the output tensor \texttt{a} of $k_1$.
$k_4$ reads its inputs from the output tensor \texttt{f} of $k_3$ and writes to the input tensor \texttt{j} of $k_5$.
VTC achieves only modest speed-up ($\sim 1.1\times$) and memory saving ($\sim 4.4\%$) on this model because most data movement elimination (e.g., optimizations related to \texttt{f}, \texttt{g}, \texttt{h}, \texttt{i} and \texttt{j}) can already be achieved by operator fusion in the baselines.

\paragraph{YOLOv11.}
~\Cref{fig:yolo} shows the computation graph of C3K2 Block, a recurring component in YOLOv11.
This block contains two data movement operators, \texttt{Split} and \texttt{Concat}.
Traditional frameworks like TensorRT and PyTorch execute \texttt{Split} in an individual kernel and \texttt{Concat} in another kernel fused with \texttt{Swish} and \texttt{Add} (taking \texttt{b}, \texttt{c}, \texttt{d} as inputs and \texttt{Y} as the output).
In contrast, \sys eliminates both data movement operators by making \texttt{a}, \texttt{b}, \texttt{c} and \texttt{e} virtual tensors of \texttt{Y}.
Although \texttt{Split} and \texttt{Concat} are not fully contiguous at batch size 16 (\Cref{fig:split}), each contiguous chunk ($16\times 160\times 160$ elements) is large enough to read and write in a coalesced manner, preserving high memory I/O bandwidth.
Moreover, these two data movement operators constitute the primary memory footprint bottlenecks, as they operate on large tensors (\texttt{a}, \texttt{b}, \texttt{c}, \texttt{e} and \texttt{Y}).
After \sys's optimization, all these tensors except \texttt{Y} become virtual, eliminating the need for physical memory allocation and reducing peak memory consumption.

\section{Related Work}
\paragraph{Virtual tensors.}
~VTensor~\cite{xue2023vtensor} is a programming framework which decouples tensor layouts from the programming interface, enabling developers to write layout-agnostic operations which significantly reduces lines of code.
Separately, vLLM~\cite{kwon2023efficient} and vTensor~\cite{xu2024vtensor} are virtual memory management systems \emph{specifically} designed for LLM serving.
Unlike these systems, \sys optimizes \emph{general} DNNs including LLMs with a fundamentally different virtual tensor notation.

\paragraph{Layout optimizations.}

~TensorRT~\cite{tensorrt} explores various layout options for 4D tensors and automatically applies the necessary layout transformations to determine the optimal layout.
SmartMem~\cite{niu2024smartmem} studies layout transformation elimination based on predefined rules and develops efficient memory layouts for 2.5D memory on mobile devices.
In contrast to these layout optimizations, \sys proposes a more general optimization for data movements.
Instead of focusing solely on layout operators (mostly \texttt{Reshape} and \texttt{Transpose}), \sys's optimization can be applied to all data movement operators.

Furthermore, \sys's points-to graph construction algorithm automatically explores virtual tensor creation strategies, thereby expanding the optimization space beyond rule-based approaches.

\paragraph{Tensor graph optimizations.}

~TensorFlow~\cite{abadi2016tensorflow} and TensorRT~\cite{tensorrt} leverage manually designed rules to perform graph-level transformations.
Recently, automated graph optimizers~\cite{jia2019taso,tensat,wang2021pet,zheng2023einnet} have emerged to automatically generate subgraph substitutions.

\sys's optimization is orthogonal to these graph optimizations and can be used as an independent optimization pass after graph optimization in ML compilers.

\paragraph{Kernel orchestration} is the process of mapping a computation graph to hardware-specific kernels~\cite{hu2024optimal}.
This problem is commonly addressed by operator fusion~\cite{chen2018tvm,dnnfusion,abadi2016tensorflow,tensorrt}, which fuses multiple operators into a single kernel, thereby eliminating data movements of intermediate results.

However, this type of work cannot optimize data movement operators that are hard to fuse with other operators.
\sys takes a computation graph after operator fusion as input and further eliminates data movements between fused kernels.

\paragraph{Kernel generation.}
~Numerous ML compilers~\cite{chen2018tvm,zheng2020flextensor,zheng2020ansor} automatically generate hardware-specific kernels for DNN computation based on Halide's algorithm-schedule separation~\cite{ragan2013halide}.
Rammer~\cite{ma2020rammer} and Hidet~\cite{ding2023hidet} approach kernel generation by utilizing a task abstraction of workloads.
Mirage~\cite{wu2025mirage} employs superoptimization techniques on a multi-level compute hierarchy to generate efficient GPU kernels.
As \sys only modifies the global memory I/O (stages 1 and 3) of a hardware-specific kernel, it can be easily integrated with virtual tensor support in these kernel generation frameworks.

\paragraph{Hand-optimized kernels} crafted by domain experts can achieve state-of-the-art performance on specific DNN workloads.
For instance, FlashAttention~\cite{flashdecoding,dao2022flashattention} and FlashInfer~\cite{flashinfer} design highly optimized GPU kernels for self-attention calculation.
\sys can also incorporate these manually optimized kernels by adding virtual tensor support.

\section{Future Work}

While our current virtual tensor construction algorithm provides performance guarantees through iterative profiling, this approach can incur substantial compilation overhead for time-sensitive applications. The VTOG formulation we present opens opportunities for developing more efficient algorithms that can compute high-quality virtual tensor strategies with reduced profiling requirements.

Our current implementation is based on Triton. However, Triton exhibits known performance limitations on newer GPU architectures, particularly NVIDIA Blackwell GPUs, where domain-specific languages such as CUTLASS demonstrate superior performance.
Relatedly, the virtual tensor design does not explicitly consider emerging hardware-specific memory transfer mechanisms such as NVIDIA's Tensor Memory Accelerator (TMA), a hardware unit for asynchronous multi-dimensional memory transfers.
While Triton handles some TMA patterns implicitly, explicit TMA awareness in the virtual tensor design could open additional optimization opportunities.
Extending virtual tensor support to alternative DSLs and compilation frameworks, together with native handling of such bulk transfer mechanisms, is a promising direction for further performance gains on newer hardware architectures.

\section{Conclusion}

\noindent Data movement optimizations have become extremely important in tensor compilers to achieve good performance. \sys is the first compilation framework to eliminate all unnecessary data movement that spans across all data movement operators in a given tensor computational graph. We introduce the concept of a virtual tensor that allows \sys to track data movement between different physical tensors without instantiating intermediate tensors. \sys's data movement elimination algorithm allows us to achieve superior performance on a number of different DNN models including appreciable gains on large language model inference.

\section*{Acknowledgments}
We thank Zhihao Jia, Hao Guo, Yuhao Ge and Yueying Li for their feedback and help on this work.
This work was supported by PRISM and ACE, two of the seven centers in JUMP 2.0, a Semiconductor Research Corporation (SRC) program sponsored by DARPA, by NSF under the grant CCF-2316233, and by generous gifts from Qualcomm.

\bibliographystyle{plain}

\end{document}